\documentclass[12pt]{article}
\usepackage[T1]{fontenc}
\usepackage{titlesec}
\usepackage{lmodern}

\titleformat*{\section}{\Large\bfseries\sffamily}
\titleformat*{\subsection}{\large\bfseries\sffamily}

\usepackage{amsmath}
\usepackage{listings}
\usepackage{graphicx}
\usepackage{hyperref}

\begin{document}

\fontfamily{cmss}\selectfont

\title{An Exact Bitwise Reversible Integrator}
\author{Jos Stam \\
NVIDIA}

\maketitle

\begin{abstract}
   At a fundamental level most physical equations are time reversible. In this paper we propose an integrator that preserves this property at the discrete computational level. Our simulations can be run forward and backwards and trace the same path exactly bitwise. We achieve this by implementing theoretically reversible integrators using a mix of fixed and floating point arithmetic. Our main application is in efficiently implementing the reverse step in the adjoint method used in optimization. Our integrator has applications in differential simulations and machine learning (backpropagation).
   
 \end{abstract}

\def\cL{\mathcal{L}}
\def\bfa{\mathbf{a}}
\def\bff{\mathbf{f}}
\def\bfq{\mathbf{q}}
\def\bfp{\mathbf{p}}
\def\bfz{\mathbf{z}}

\def\baq{\hat{\bfq}}
\def\bap{\hat{\bfp}}

\def\pd{\partial}

\section{Introduction}

Most fundamental physical equations are time reversible. Simulations of these equations running forward in time are indistinguishable from those running in reverse. However, our intuition tells us that this is not the case in real life. Rarely do we see bits of shards of broken glass reverse to form a bottle resting on a table. However, we are familiar with bottles falling off a table and smashing on the floor into many shards of glass. This is the so called ``paradox of the arrow of time.'' How can time reversible fundamental equations give rise to seemingly irreversible phenomena? We will not dwell on these philosophical considerations, see for example \cite{ArrowOfTime}. An easy way out of this conundrum is to consider that all phenomena satisfying the reversible physical laws are possible but that some are very improbable. The ones with high probability to occur are the ones we observe in practice.

Examples of reversible processes include molecular dynamics, the Euler equations of fluid motion, planetary systems, and so forth. Although these systems are ideal cases they are nonetheless ubiquitous in simulation. Or at least until dissipative effects are introduced.

In computers, physical processes are discretized and integrated using a time step $h$. The state of the system (particle positions and velocities for example) is some finite quantity $S$. Both $h$ and $S$ are \textbf{not} represented as idealized (mathematical) real numbers but using finite representations. Basically bits, strings of ones and zeros. For example a time step of $h=\pi$ is represented in 32 bit IEEE 754 float standard as the following string of bits: \[0\;10000000\;10010010000111111011011,\]
or alternatively in hex notation: {40420FDB} \cite{IEEE754}. Below we will show another way of discretizing reals like $\pi$ using fixed point numbers.

In this paper, we discuss a class of reversible integrators for Hamiltonian systems. The most common example are Newton's equations with conservative forces that are the gradient of a potential field. When using floating point numbers these integrators are only approximately reversible. By using fixed numbers which basically amounts to integrating with integers, we obtain exact bitwise reversible simulations. This means that if we integrate the equations forward for an arbitrary number of time steps $h$ and then decide to integrate backwards with $-h$ we recover any previous state exactly at a bitwise level. Like $1200045 = 1200045$ not approximately like $1200045 \approx 1200044$.

Many applications can benefit from reversible integrators. Here we list a few.

\begin{itemize}
\item Animators can browse back and forth in time to refine a simulation's results. This is akin to how traditional animators flip through their sequence of drawings using a {\em flipbook}. 
 
\item In Machine Learning and Optimization of physical systems a very popular method is the {\em adjoint method} \cite{Lagrange,BackProp,Giles2000}. This technique requires the reverse simulation of the physical system. Usually this is achieved by storing the results of the forward solution. We will discuss this application in more length below.

\item Reversible simulations do not suffer from energy drift and are therefore relatively stable for long term simulations. The energy is not exactly conserved but bounded. They are however, not unconditionally stable and have an upper bound on the time step $h$.

\end{itemize}

This paper was mainly inspired by the seminal work of Levesque and Verlet \cite{Levesque93}. This works fits into the class of {\em Geometrical Integrators} that preserve properties of the underlying equations they simulate \cite{Hairer2002}. We are of course not the first ones to use this technique. Rein and Tamayo apply it to simulating n-body systems like our solar system \cite{JANUS}. This technique has also been applied to analyzing complex statistical mechanical systems \cite{Hoover}. In this paper we clearly present how to implement this technique, providing source code. We also provide a novel application in computing gradients in the context of reversible neural ODEs.

\section{Hamiltonian Systems}

\subsection{A Simple Spring}

One of the simplest examples of a Hamiltonian system is that of a one-dimensional spring. Not just any spring but one that has unit mass and unit stiffness. The spring's state is then defined at any point in time by two variables its position $q(t)$ and its velocity (momentum) $p(t)$. The Hamiltonian (energy) in this case is equal to the sum of kinetic and potential energies: $H(q,p)=\frac{1}{2}p^2+\frac{1}{2}q^2$. Energy conservation then directly implies that a solution lies on a circle in phase space $q-p$ that contains the initial condition $(q(0),p(0))^T$. We know that the spring satisfies Newton's second order differential equation: $\ddot{q} = -q$. Alternatively, the equations can be written as a first order differential equation using the Hamiltonian:
\begin{equation}
\dot{q} = \frac{\pd H}{\pd p}
\;\;\; \mathrm{and} \;\;\;
\dot{p} = -\frac{\pd H}{\pd q}.
\label{eq:Hamilton}
\end{equation}
For our specific Hamiltonian we have that $\frac{\pd H}{\pd q} = q$ and $\frac{\pd H}{\pd p} = p$. Introducing the complex number $z(t) = q(t)+ip(t)$, Eq. \ref{eq:Hamilton} becomes $\dot{z}=-iz$ and its solution is a rotation of the initial state
\begin{equation}
    z(t)=e^{-it}\;z(0) \label{eq:spring}
\end{equation} 
and lies on a circle. Obviously, the solution is exactly reversible.

\subsection{General Case}

The general case is multi-dimensional. Our state consists of $n$ {\em generalized coordinates} $\bfq$ and $n$ {\em generalized momenta} $\bfp$. We assume that the Hamiltonian depends solely on the state $H(\bfq,\bfp)$ satisfying Eq. \ref{eq:Hamilton} with $q$ and $p$ in boldface. Consider any property $A(\bfq,\bfp)$, its evolution over time is implicitly given by its dependence on the state $(\bfq,\bfp)^T$. Indeed, using the chain rule of differentiation and Eq. \ref{eq:Hamilton}, we get that:
\begin{equation} 
\dot{A} = \dot{\bfq} \frac{\pd A}{\pd \bfq} +  \dot{\bfp} \frac{\pd A}{\pd \bfp} = 
\frac{\pd H}{\pd \bfp} \frac{\pd A}{\pd \bfq} -
\frac{\pd H}{\pd \bfq} \frac{\pd A}{\pd \bfp} =
i\cL\;A, \label{eq:Liouville}
\end{equation}
where
\begin{equation}
    i\cL = \frac{\pd H}{\pd \bfp} \frac{\pd}{\pd \bfq} -
    \frac{\pd H}{\pd \bfq} \frac{\pd}{\pd \bfp} =
    i\cL_q + i\cL_p.
    \label{eq:Liouville_def}
\end{equation}
is called the {\em Liouville operator}. Note that by convention the imaginary number $i$ is there to emphasize that this operator is Hermitian. Eq. \ref{eq:Liouville} can formally be solved as follows:
\begin{equation}
A(t) = e^{i\cL t} \; A(0). \label{eq:general}
\end{equation}
Similarly to the simple spring example, we can introduce a $2n$-dimensional variable $\bfz=(\bfq,\bfp)^T$ and plug it into Eq. \ref{eq:general} for each of its components  ($A=z_i$) to get an equation for the evolution of our state:
\begin{equation}
    \bfz(t) = e^{i\cL t} \; \bfz(0). \label{eq:evolution}
\end{equation}
We can regard this equation as a "rotation" (unitary  transformation) of the initial state over time. The evolution is therefore exactly reversible and preserves energy analogously to the one dimensional spring case above. We can verify that this is indeed a special case $i\cL z=p\frac{\pd z}{\pd q}-q\frac{\pd z}{\pd p} = p - iq = -i(q+ip) = -iz$.

\subsection{Discretization}

We now derive a reversible discrete integrator from the theoretical considerations. We follow the work of Tuckerman {\em et al.} \cite{Tuckerman92}. The idea is to discretize Eq. \ref{eq:evolution} by replacing the Liouvillian by its decomposition (Eq. \ref{eq:Liouville_def}) and replace the continuous time variable $t$ with its discrete counterpart $h$.
\begin{equation}
    \bfz(h) = e^{(i\cL_q+i\cL_p)h} \; \bfz(0) = 
    e^{i\cL_q h} \; e^{i\cL_p h} \; \bfz(0) + O(h^2).
    \label{eq:discrete_Liouville}
\end{equation}
The last equality is exact only when $i\cL_q$ and $i\cL_p$ commute which in general is not the case. The propagation operator on the left hand side is also non reversible. The clever idea is to replace the propagator with an equivalent symmetrical one:
\begin{equation}
    \bfz(h) = e^{i \cL_q \; h/2} \; e^{i \cL_p \; h} \; e^{i \cL_q \; h/2} \; \bfz(0) + O(h^3).
    \label{eq:discrete_integrator}
\end{equation}
This integrator is unitary and invertible as desired and of order two. This is sufficient for our bitwise-accurate integrator below. However, higher order integrators can be constructed similarly if needed \cite{Hairer2002}.

Now using first order approximations of the integrators in Eq. \ref{eq:discrete_integrator}
becomes the well known {\em Position Verlet} integration step:
\begin{eqnarray*}
    \bfq(h/2) &=& \bfq(0) + h/2 \; \frac{\pd \hat{H}(0)}{\pd \bfp} \\
    \bfp(h) &=& \bfp(0) + h \; \left( - \frac{\pd \hat{H}(h/2)}{\pd \bfq} \right) \\
    \bfq(h) &=& \bfq(h/2) + h/2 \; \frac{\pd \hat{H}(h)}{\pd \bfp},
\end{eqnarray*}
where $\hat{H}(t)=H(\bfq(t),\bfp(t))$. We point out that by exchanging the roles of $\bfq$ and $\bfp$ in Eq. \ref{eq:discrete_integrator} we get the {\em Velocity Verlet} integrator which is also time reversible. This method is also more commonly referred to as {\em Leap Frog}.

In most application the Hamiltonian is the sum of kinetic energy and a potential $V$: $H=\frac{1}{2}|\bfp|^2 + V(\bfq)$. In this case we can introduce the conservative force $\bff(t)=-\frac{\pd V(\bfq(t))}{\pd \bfq}$ which results in the following Position Verlet integrator:
\begin{eqnarray}
    \bfq(h/2) &=& \bfq(0) + h/2 \; \bfp(0) \nonumber \\
    \bfp(h) &=& \bfp(0) + h \; \bff(h/2) \label{eq:posVerlet}\\
    \bfq(h) &=& \bfq(h/2) + h/2 \; \bfp(h). \nonumber
\end{eqnarray}

This Position Verlet formulation will be our go to reversible integrator in the rest of the paper. We note that these integrators are not unconditionally stable. A strict stability bound is hard to establish because the forces can be non-linear. A linear stability analysis however suggest the bound $h\omega<1$, where $\omega$ is the highest ``frequency'' of the linear system. Before we present our implementation we discuss how real numbers are represented in computers.

\section{Float versus Fixed}

Once we have our reversible integrator our job seems to be done. Usually we implement the integrator in some programming language that has some built in floating point data type. We run the simulation forward in time for $N$ steps and then run it backwards ($h \leftarrow -h$) for $N$ steps. Ideally, we should return to the initial state exactly. This is usually not the case. As James Gosling the creator of {\em Java} once famously said ``95\% of folks out there are completely clueless about floating-point.'' Floats are tricky. They do not behave like the ideal real numbers they try to represent. They have the advantage that they can handle a wide range of magnitudes as they are based on scientific notation. However, the basic properties of real numbers are often not satisfied, for example $0.1 + 0.2 \neq 0.3$. Integers do not suffer from this problem as they are discrete.

Enter the {\em Fixed Point} numbers. This format was popular before floats were implemented in hardware. They work well if your data is bounded. Since our integrators are bounded for reasonable time step sizes $h$ we can assume without loss of generality that our reals lie in the interval $[1,-1]$. We can then approximate a real number $R$ by the following integer $I$:
\begin{equation}
    I = \lfloor R \; \times \; \mathrm{BIG\_INT} \rfloor,
    \label{eq:pfixed}
\end{equation}
where BIG\_INT is a large integer and $\lfloor\cdot\rfloor$ is the integer part of a real number.
The only operation needed in our integration algorithm is addition as shown below in the implementation. To get a real number from a fixed number, we simply ``invert'' Eq. \ref{eq:pfixed}:
\begin{equation*}
    R = I \; / \; \mathrm{BIG\_INT}.
\end{equation*}
Actually, our implementation is hybrid and uses floating points to compute forces and fixed points when integrating.

\section{Implementation}
 
In the following C++ code only the integration steps use the fixed format. Both the coordinates and the velocities are represented in fixed format. The time step can be any arbitrary float. The simulation is completely controlled by the force function which uses only floats. Existing simulations can therefore easily be implemented in this framework. We represent our fixed points as 64 bit integers ``int64\_t'', a type standardized since C++11.
 
 {\small
 \begin{lstlisting}
typedef int64_t fixed;
const fixed c_max_fixed = 0x1000000000000000;
    
static float x2f(const fixed i_f){
   return i_f / (float)c_max_fixed;
}
static fixed f2x(const float i_f) {
   return (fixed)(i_f * c_max_fixed);
}

static fixed * x, * v;
static int size;
// define your favorite force field here.
static float force(const int i, const float i_x); 

void integrate(const float h) {
   const float h2 = 0.5f * h;
    
   for (int i = 0; i < size; i++) {
      x[i] += f2x( h2 * x2f(v[i]) );
   }
   for (int i = 0; i < size; i++) {
      float f = force( i, x2f(x[i]) );
      v[i] += f2x( h * f );
   }
   for (int i = 0; i < size; i++) {
      x[i] += f2x( h2 * x2f(v[i]) );
   }
}
\end{lstlisting}
}

We also implemented the integrator in {\em JavaScript (JS)} to create a web based demo that is easily shared. Ironically, we were faced with the problem that JS does not differentiate between integers and float numbers. It only has the monolithic {\em Number} type. We found a solution by treating fixed point numbers as strings. We lost some efficiency of course. The implementation is as follows:
{\small
\begin{lstlisting}
var XN = 1000000000;
function X2F(x) { return (+x) / XN; }
function F2X(f) { ((f*XN)|0).toString(); }
function XADD(x1, x2) { (+x1 + +x2).toString(); }
\end{lstlisting}
}
The code is a bit obfuscated but JS programmers will recognize the hack to get an integer from a Number by taking the bitwise ``or'' with zero. The ``+'' operator transforms a string into a Number. See the accompanying {\em HTML} file from the link given below for the entire implementation. We would like to hear from readers who have a better solution. Note that this is just a quirk of the JS language. For most typed languages, like in our C++ implementation above, there is no performance hit. 

\begin{figure}
\includegraphics[scale=0.8]{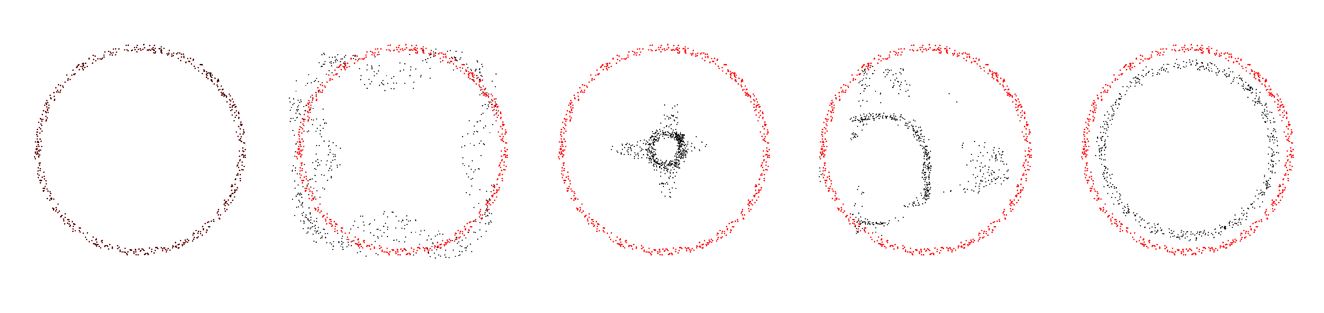}
\caption{Gravitational collapse of a ring and back.}
\label{fig:ring}
\end{figure}

\section{Results}

We implemented the exact bitwise integrator for various forces. In each example we run the simulation forward in time and then reverse the time step $h$ and observe that the initial state is exactly recovered. Figure \ref{fig:ring} shows 5 frames of a particle system which evolves under pair-wise gravitational-like forces. The initial ring configuration first collapses and after a time reversal exactly returns to its initial state.

We also implemented a chaotic pendulum simulation in JavaScript so it can easily be shared on the internet. You can find the demo on the following web page: \href{http://www.josstam.com/reversible}{josstam.com/reversible}. The html source is self-contained and includes all the JS code. The readers should feel free to experiment with the code.

\section{Application: Reversible Adjoint / Neural ODEs}

Reversible integrators can play an important role in optimization and machine learning. First, we have to introduce some ideas from optimization. We provide a very concise derivation of the so called {\em adjoint method} and then apply it to our Hamiltonian framework.

The goal in optimization and machine learning is to find an optimal set of controls/weights $\theta$ that minimize/maximize some {\em cost} function $J$. The function to optimize usually depends on a state $u(\theta)$ which satisfies some equation/constraint $E(u,\theta)$. This is a constrained optimization problem.
\begin{equation}
    \text{Find:} \;\; \theta^* = \text{argmin}_{\theta} \; \; {J(u(\theta))} \;\;\; \text{such that} \;\;\; E(u,\theta)=0.
    \label{eq:Optim}
\end{equation}
Two important examples follow.
(1) A Vanilla {\em Neural Network} computes an output $y=f(x,\theta)$ from a set of node values $x$ and weights $\theta$, the state is $u=(x,y)$, the constraint is $E=-y+f(x,\theta)=0$ and $J$ is the cost function.
(2) An {\em Ordinary Differential Equation (ODE)} also fits this framework. In this case the state $u(t,\theta)$ is some continuous quantity that evolves over time and depends on continuous controls $\theta(t)$ like external forces. The function $J$ models some goal we want to achieve, like matching a keyframe. The state $u$ satisfies an ODE: $E=-\dot{u} + f(u,\theta)$. The ``Neural ODEs'' paper combines these two examples in a clever way \cite{NeuralODE}.

Equation \ref{eq:Optim} can be solved for the controls $\theta$ by many optimization techniques \cite{Optimization}. Usually they involve the gradient of the cost function with respect to the controls.
\begin{equation}
    \delta J := \frac{dJ}{d\theta} = \frac{\pd J}{\pd u} \; \frac{d u}{d\theta} =: J_u \; \delta u.
    \label{eq:dJ}
\end{equation}
The gradient depends on the differential $\delta u$ for which we can derive an equation by differentiating $E=0$:
\begin{equation}
    0=\delta E = \frac{d E}{d \theta} = \frac{\pd E}{\pd u} \frac{d u}{d \theta} + \frac{\pd E}{\pd \theta} \;\;\; \text{or} \;\; \; E_u \; \delta u = - E_\theta.
    \label{eq:forward}
\end{equation}
This is a linear equation for $\delta u$, whose solution we can then plug into Eq. \ref{eq:dJ} to get our desired $\delta J$. This equation becomes expensive to solve in the presence of many controls like in deep learning which have many weights.

Fortunately, there is an alternative known as the {\em adjoint method}. Instead of solving \ref{eq:forward} we solve an {\em adjoint} equation involving a new variable $a$, known as a {\em Lagrange multiplier} or {\em adjoint variable}.
\begin{equation}
    (E_u)^* \; a = -J_u^* \;\;\; \text{and}\;\;\;
    \delta J^\prime = a^* E_\theta.
    \label{eq:reverse}
\end{equation}
Where ``$^*$'' is the adjoint operation (transpose for vectors and matrices). The adjoint equation is independent of the controls. The fact that $\delta J^\prime = \delta J$ can be established in a one liner proof: 
\begin{equation}
\delta J = J_u \delta u = -\left(E_u^* a\right)^* \delta u = -a^* E_u \delta u = a^* E_\theta = \delta J^\prime.
\end{equation}
For the case of an ODE ($d/dt^*=-d/dt$) we get in agreement with \cite{NeuralODE}:
\begin{equation*}
    -\dot{a} = \left(\frac{\pd f(u,\theta)}{\pd u}\right)^* \; a + \left(\frac{\pd J}{\pd u}\right)^*.
\end{equation*}
This is a linear ODE that runs in reverse. It depends on the values of $u$ in the forward simulation. Usually these states are stored like in neural networks or dealt with using clever check-pointing techniques. For our reversable integrators we do not have that problem since we can retrace our path along with the adjoint equation. More concretely, let $\bfa = (\baq,\bap)^T$ be the adjoint state to our coordinate/momentum coordinates. Then the adjoint Position Verlet (Eq. \ref{eq:posVerlet}) is:
\begin{eqnarray}
    \bap(-h/2) &=& \bap(-h) - h/2 \; \baq(-h) \nonumber \\
    \baq(0) &=& \baq(-h) - h \; \left(\frac{\pd \bff}{\pd \bfq}(-h/2)\right)^* \; \bap(-h/2) \label{eq:adjPosVerlet}\\
    \bap(0) &=& \bap(-h/2) - h/2 \; \baq(0). \nonumber
\end{eqnarray}
Interestingly, this is a Velocity Verlet integration for the adjoint. Similarly, if we had started with a Velocity Verlet integration we would have obtained a Position Verlet integration scheme for the adjoint.

\begin{figure}
\includegraphics[scale=0.55]{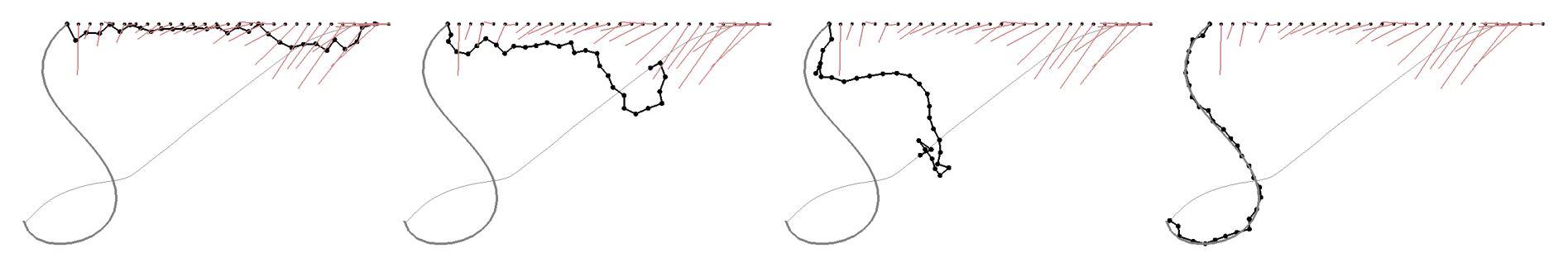}
\caption{Chain controlled to hit the "S" target.}
\label{fig:control}
\end{figure}

We used this framework to keyframe control a chaotic pendulum to match a keyframe similarly to \cite{adjointFluids,ParticleAdjoint}. Our controls are the initial velocities of the chain. A sequence of frames is shown in Figure \ref{fig:control}. The controls are shown in red.

\section{Conclusion and Future Work}

In this paper we have introduced an exact reversible integrator for Hamiltonian systems. Hopefully, we have given enough theoretical background and actual code to make this technique understandable and useful. We also showed how to use this solver to efficiently solve a class of optimization problems using the adjoint method.

In future work we would like to explore applications in machine learning, e.g. reversible neural networks. Also we would like to extend this integrator to non-Hamiltonian reversible systems. Another challenge is to derive an exact reversible integrator for the Euler equations of fluid dynamics.

In general we hope to inspire people to explore the interchange between techniques in machine learning and simulation.

\bibliographystyle{plain}
\bibliography{bibsample}       

\end{document}